\documentclass[aps,showpacs,showkeys,superscriptaddress]{revtex4}

\usepackage{graphicx}
\usepackage{makeidx}
\usepackage[]{hyperref}
\usepackage{amsmath}

\usepackage{amssymb,epsfig,graphics}
\usepackage{amscd}
\usepackage{verbatim}

\usepackage{amsmath}
\usepackage{amsthm}
\usepackage{amsfonts}
\usepackage{dsfont}

\begin{document}
\title{Boundary conditions and Green function approach of the spin-orbit interaction in the graphitic nanocone}

\author{J. Smotlacha}\email{smota@centrum.cz}
\affiliation{Bogoliubov Laboratory of Theoretical Physics, Joint
Institute for Nuclear Research, 141980 Dubna, Moscow region, Russia}
\affiliation{Faculty of Nuclear Sciences and Physical Engineering, Czech Technical University, Brehova 7, 110 00 Prague,
Czech Republic}

\author{R. Pincak}\email{pincak@saske.sk}
\affiliation{Institute of Experimental Physics, Slovak Academy of Sciences,
Watsonova 47,043 53 Kosice, Slovak Republic}
\affiliation{Bogoliubov Laboratory of Theoretical Physics, Joint
Institute for Nuclear Research, 141980 Dubna, Moscow region, Russia}

\date{\today}

\begin{abstract}
The boundary effects affecting the Hamiltonian for the nanocone with curvature--induced spin--orbit
coupling were considered and the corresponding electronic structure was calculated. These boundary effects include the spin--orbit coupling, the electron mass acquisition and the Coulomb interaction. Different numbers of the pentagonal defects in the tip were considered. The matrix and analytical form of the Green function approach was used for the verification of our results and the increase of their precision in the case of the spin--orbit coupling.
\end{abstract}

\pacs{ 73.22.Pr, 81.05.ue, 71.70.Ej, – 72.25.-b.}

\keywords{Tight-binding method,
Graphitic nanocone, Spin--orbit coupling, Boundary conditions, Coulomb interaction, Green function approach}

\maketitle

\section{Introduction}\

The electronic structure of the carbon nanoparticles is of great interest in the todays physical research. These nanostructures have a large potential use as electronic nanodevices in the computer science. The main structure of this kind is the plain graphene, but other kinds of the nanostructures like fullerene, nanotubes etc. were already largely investigated.

The main characteristics of the electronic properties is the local density of states ($LDoS$). It can be calculated using the different methods; here we use the continuum gauge field-theory approach \cite{mele, callaway} which leads to the solution of a 2-component Dirac-like equation. The result we get is the sum of the squares of the components of the normalized solution. This method was used in \cite{kochetov, pincak, smotlacha, ando, sitenko, herrero1, herrero2, wormhole} for the calculation of the electronic structure in the close surrounding of different kinds of the defects and of fullerene, nanocone, nanotubes and wormhole.

One phenomenon has not been yet considered enough in the graphene nanostructures: the effect of the spin-orbit coupling ($SOC$). The reason is that this effect is very weak in the case of the plain graphene, but it can grow considerably in other kinds of the carbon nanostructures because of the curvature and the orbital overlap of the next-nearest neighbouring atomic sites. Furthermore, the influence of the impurities could play a crucial rule. Then, we distinguish the intrinsic and the extrinsic (Rashba) term of the $SOC$. A lot of theoretical work concerning this effect was performed in \cite{kane, ando, Kuemmeth, Fang, Steele, Brataas, Jeong, Valle, PPN}.

There are more ways how to incorporate these terms into the Hamiltonian for the ground state of the given nanostructure. For the case of the graphitic nanocone \cite{sitenko}, different methods are described in \cite{pincak} and \cite{choudhari}: in these papers, the terms corresponding to the $SOC$ are added into different blocks of the corresponding matrix and, consequently, it is mixed with different terms of the remaining part of the Hamiltonian. Here, we will be more interested in the method used in \cite{pincak}, where the used formalism is an analog of the procedure used for the case of the nanotubes \cite{ando}.

In \cite{pincak}, the boundary effects were not considered. Their significance is (among others) connected with the needs of the quadratic integrability of the resulting wave function. They come from the rapid change of the geometry close to the conical tip and from the finite size of the conical nanostructure. The change of the geometry could give rise to the mass of the fermion \cite{wormhole}, but it could be also simulated by an addition of a fictious charge into the conical tip \cite{chakraborty}. Next, the influence of the pseudopotential is also significant \cite{pincak1}.

Finding of the analytical solution of the resulting Hamiltonians is a very difficult task. In most of our calculations, we use the numerical methods. To verify them and achieve a higher precision, the Green function approach is also used.

In this paper, we review the results of the calculations concerning the influence of the $SOC$ on the electronic structure of the graphitic nanocone in \cite{pincak}, discuss the influence of the boundary effects and incorporate
 the Green function formalism. In section II, we remind the main results from \cite{pincak}. In section III, we present the influence
of the possible boundary effects connected with the topology of the molecular surface: the relativistic acquisition of the mass connected with the curvature and the Coulomb interaction. Finally, in section IV, we verify some our numerical results using the Green function method and we derive the perturbation scheme to the second order for the purpose of getting more precise results of the calculation of the $LDoS$.\\

\section{Local density of states of the nanocone influenced by the spin--orbit coupling}\

In \cite{pincak}, we derived the Hamiltonian for the curvature--induced $SOC$ in the graphitic nanocone using the method of Ando \cite{ando}. We started with the case of the Hamiltonian without the $SOC$ \cite{sitenko} and after incorporating the appropriate $SOC$ terms, we finally got
\begin{equation}\label{HamSOC}\hat{H_{s}}=\hbar v\left(\begin{array}{cc}0 & \partial_r-{\rm i}\frac{1}{r}\xi_x\sigma_x(\vec{r})-\frac{{\rm i}s\partial_{\varphi}}{(1-\eta)r}-\frac{A_y}{r}\sigma_y-\frac{3\eta}{2(1-\eta)r}+\frac{1}{2r}\\
-\partial_r+{\rm i}\frac{1}{r}\xi_x\sigma_x(\vec{r})-\frac{{\rm i}s\partial_{\varphi}}{(1-\eta)r}-\frac{A_y}{r}\sigma_y-\frac{3\eta}{2(1-\eta)r}-\frac{1}{2r} & 0\end{array}\right),\end{equation}
where all the parameters connected with $SOC$ and with the geometry of the nanocone are described in the mentioned paper.
To calculate the $LDoS$, we solve the equation
\begin{equation}\label{DirEq}\hat{H}_s\psi(r,\varphi)=E\psi(r,\varphi)\end{equation}
and finally get the matrix equation (9) in \cite{pincak}. There, this equation is solved in a numerical way and the resulting $LDoS$ for $n=0$ is sketched in Fig. 3 of that paper.

\begin{figure}[htbp]
\includegraphics[width=160mm]{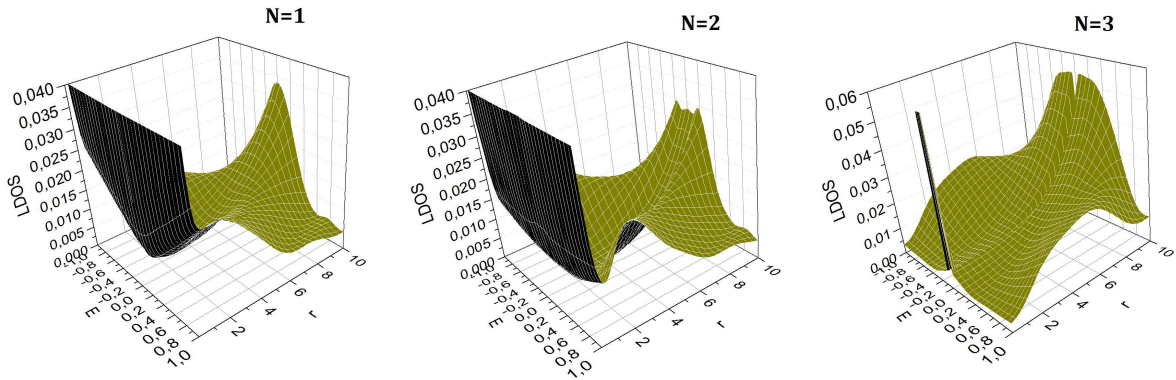}
\caption{3D graphs of the $LDoS$ for the graphitic nanocone influenced
by the $SOC$. Here, the $LDoS$ corresponds to the sum of the solutions corresponding to $n=-1,0,1,2,3$. The number of the defects in the tip in the particular cases: $N=1$ (left), $N=2$ (middle) and $N=3$ (right).}\label{LDoS3Dspin}
\end{figure}

The numerical method of finding the solution used in \cite{pincak} can be used for other modes. In Fig. \ref{LDoS3Dspin}, we see the $LDoS$ for the same numbers of the defects, but here it corresponds to the sum of the solutions corresponding  to $n=-1,0,1,2,3$. We can see that while for the case of 1 and 2 defects in the tip, the $LDoS$ should grow to infinity for $r=0$ for an arbitrary value of energy, this effect is restricted only to zero energy in the case of 3 defects in the tip.

\begin{figure}[htbp]
\includegraphics[width=160mm]{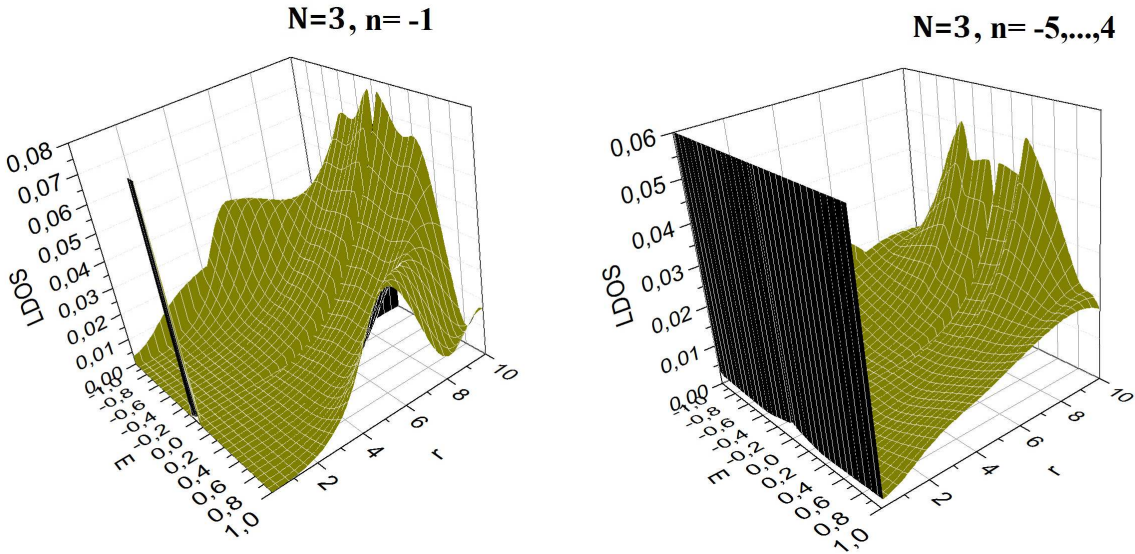}
\caption{3D graphs of the $LDoS$ for the graphitic nanocone with 3 defects in the tip influenced
by the $SOC$: $n=-1$ (left) and the case corresponding to the sum with $-5\leq n\leq 4$ (right).}\label{LDoS3Dspin3d}
\end{figure}

From these results follows that there could be a strong localization of the electrons in the tip. But we did not consider different boundary effects which could significantly influence this behavior. Our effort is to find a quadratically integrable solution which preserves the main features of the found approach. This means that for 1 and 2 defects the rise of the $LDoS$ close to $r=0$ will be suppressed or restricted to a much smaller area around $r=0$, and the peak for $r=0$ and $E=0$ in the case of 3 defects will be preserved. In the last case of 3 defects, the calculations show that the main contribution to the revealed peak is the mode $n=-1$ and that considering the next modes the resulting character of the $LDoS$ for $r=0$ will approach the behavior of the case of 1 and 2 defects (Fig. \ref{LDoS3Dspin3d}).

The Hamiltonian (\ref{HamSOC}) is not the only possibility how to express the Hamiltonian for the $SOC$: we can also use the form of the Hamiltonian where the spin--orbit coefficients and the differential operators occupy different blocks of the corresponding matrix and find the analytical solution. This possibility is outlined in \cite{choudhari}, where the Rashba term is excluded and the intrinsic terms are moved to the diagonal positions of the Hamiltonian. So the resulting Hamiltonian (Eq. (10) in \cite{choudhari}) has the form
\begin{equation}\label{HamSOC1}\hat{H_{s}}=\hbar v\left(\begin{array}{cc}\triangle_{so}\sigma'_x & -{\rm i}\partial_r-\frac{{\rm i}}{2r}-\frac{{\rm i}\nu_s}{r}\\
-{\rm i}\partial_r-\frac{{\rm i}}{2r}+\frac{{\rm i}\nu_s}{r} & -\triangle_{so}\sigma'_x(\vec{r})\end{array}\right)\end{equation}
with $\sigma'_x=\sigma_x\cos\alpha+\sigma_z\sin\alpha$ and $\sin\alpha=1-\frac{N}{6}$. Let us stress that here the intrinsic term does not depend on $r$ and the $SOC$ depends on the vortex angle only. The solution for this Hamiltonian is given by the modified Bessel function of the second kind. On the contrary, the analytical solution of (\ref{HamSOC}) was not found; we can only say that the found numerical solution \cite{pincak} is similar to the Bessel function of the first kind.

Working on the Hamiltonian (\ref{HamSOC}) investigated in \cite{pincak}, let us investigate the boundary effects to get more precise information about the electronic structure close to $r=0$.\\

\section{Influence of boundary effects: smooth geometry and Coulomb interaction}\

\begin{figure}[htbp]
\includegraphics[width=70mm]{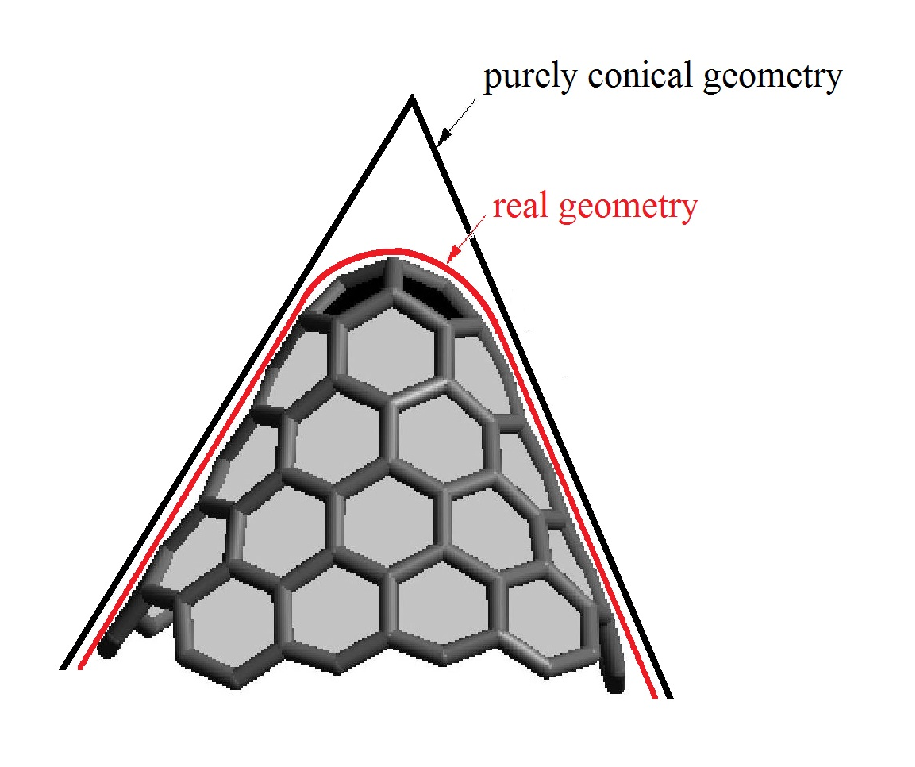}
\caption{The deviation of the geometry of the graphitic nanocone from the geometry of the real nanocone.}\label{geom}
\end{figure}

In the real graphitic nanocone, the tip has not a sharp form as one would expect for the cone geometry, but the geometry becomes smooth in this region (Fig. \ref{geom}); it means, there is such a value of $r_0$ that our predictions fail for $r<r_0$.

From the possible ways how to remove this discrepancy, we outline these 2 methods: having to find a way to simulate the smooth geometry in the conical tip, we either suppose that due to a big curvature and the connected relativistic effects, the massless fermion, for which we solve equation (\ref{DirEq}), acquires a non-negligible mass $m_0$ \cite{wormhole} or a charge is considered in the conical tip from which the Coulomb interaction comes influencing the results.

\subsection{Method of a massive fermion}\

For $r<r_0$, we can write the corrected version of the matrix equation (9) from \cite{pincak}:
\begin{equation}\label{syst1}\left(\begin{array}{cccc}m_0 & 0 & \partial_r+\frac{F}{r} &  -\frac{\rm i}{r}C\\0 & m_0 &  -\frac{\rm i}{r}D &
\partial_r+\frac{F}{r}\\ -\partial_r+\frac{F-1}{r} & \frac{\rm i}{r}D & -m_0 & 0\\
 \frac{\rm i}{r}C & -\partial_r+\frac{F-1}{r} & 0 & -m_0\end{array}\right)\left(\begin{array}{c}f_{n\uparrow}(r)\\
f_{n\downarrow}(r)\\g_{n\uparrow}(r)\\g_{n\downarrow}(r)\end{array}\right)=E\left(\begin{array}{c}f_{n\uparrow}(r)\\
f_{n\downarrow}(r)\\g_{n\uparrow}(r)\\g_{n\downarrow}(r)\end{array}\right).\end{equation}
For higher values of $r$, where the curvature is considerably smaller, the mass is considered to be zero and the form of the matrix equation is changed into the mentioned form.

We can solve this equation numerically and analytically, as well. In the case of the numerical solution, especially the case of 3 defects would cause a shift and decoupling of the peaks in the corresponding graph. However, the main purpose of using the boundary conditions - getting the solution which is quadratically integrable - is not achieved here.

For the purpose of getting the analytical solution, we suppose in (\ref{syst1}) that the mass $m_0$ acquired by the fermion is considerably larger than other effects contained in the Hamiltonian, i.e. $\xi_x=0$ and $m_0\gg r,\xi_y, F, f_{n\uparrow}$ (and the same up to the 4-th order of the powers and the differentiations). Then, the solution can be very roughly approximated as
\begin{equation}f_{n\uparrow}(r)=r^{\zeta_1}\exp\left(\zeta_2r^2\right), \zeta_1=\frac{1-6F+4F^2}{1+4F}, \zeta_2=\frac{E^2-m_0^2}{2+8F},\end{equation}
\begin{equation}f_{n\downarrow}(r)={\rm i}\frac{r^{\zeta_1}}{\xi_y}\exp\left(\zeta_2r^2\right)\cdot\left(-\zeta_1+F-1-2\zeta_2r^2-\frac{E^2-m_0^2}{2F-1}r^2\right), \end{equation}
\begin{equation}g_{n\uparrow}(r)=\frac{E-m_0}{2F-1}r^{\zeta_1+1}\exp\left(\zeta_2r^2\right),\end{equation}
\begin{equation}g_{n\downarrow}(r)={\rm i}\frac{r^{\zeta_1+1}}{\xi_y}\exp\left(\zeta_2r^2\right)\cdot\left[\frac{E-m_0}{2F-1}(\zeta_1+2-F+2\zeta_2r^2)\right].\end{equation}
The graphs of the $LDoS$ corresponding to this solution coincide with the graphs for the numerical solution with $m_0\gg E$, so the potential peaks for an arbitrary number of the defects disappear here.\\

\subsection{Method of a charge simulation}\

\begin{figure}[htbp]
\includegraphics[width=170mm]{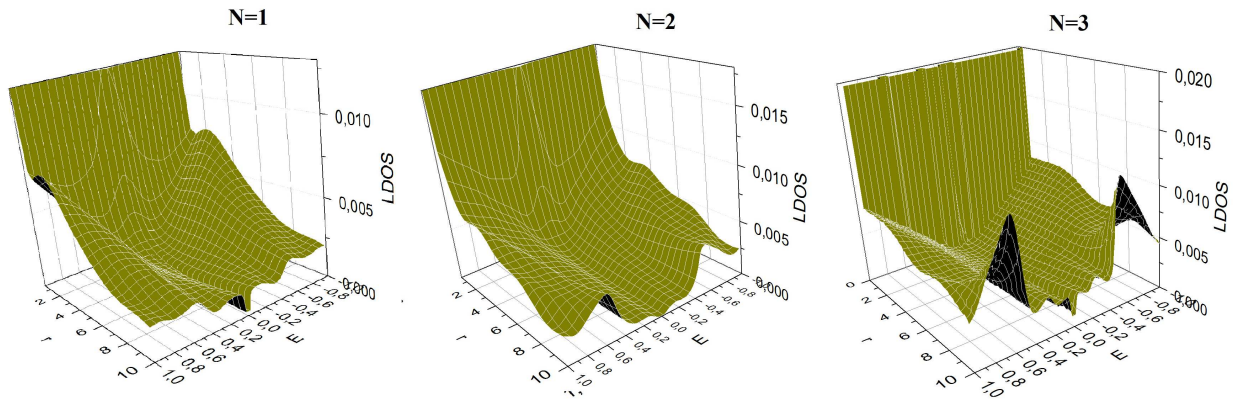}
\caption{Graphs of the $LDoS$
for the graphitic nanocone influenced by the Coulomb interaction (with the exclusion of the $SOC$) for different
distances $r$ from the tip, $-2.5\leq n\leq 2.5$ and for 1, 2 and 3 defects. We see that similarly as in the case of the $SOC$ only, we get the localization of the electrons in the tip for 1 and 2 defects at all energies but, furthermore, there is no energy restriction for the case of 3 defects as well.}\label{graph-coulomb-bezS}
\end{figure}

The influence of the charge considered in the conical tip is expressed in the Hamiltonian by the presence of the diagonal term $-\frac{\kappa}{r}$, where $\kappa=1/137$ is the fine structure constant. So this term substitutes the mass term in (\ref{syst1}):
\begin{equation}\label{syst2}\left(\begin{array}{cccc}-\frac{\kappa}{r} & 0 & \partial_r+\frac{F}{r} & -\frac{\rm i}{r}C\\0 & -\frac{\kappa}{r} & -\frac{\rm i}{r}D &
\partial_r+\frac{F}{r}\\ -\partial_r+\frac{F-1}{r} & \frac{\rm i}{r}D & -\frac{\kappa}{r} & 0\\
\frac{\rm i}{r}C & -\partial_r+\frac{F-1}{r} & 0 & -\frac{\kappa}{r}\end{array}\right)\left(\begin{array}{c}f_{n\uparrow,C}(r)\\
f_{n\downarrow,C}(r)\\g_{n\uparrow,C}(r)\\g_{n\downarrow,C}(r)\end{array}\right)=E\left(\begin{array}{c}f_{n\uparrow,C}(r)\\
f_{n\downarrow,C}(r)\\g_{n\uparrow,C}(r)\\g_{n\downarrow,C}(r)\end{array}\right).\end{equation}
Here, the coefficients $C, D$ can be considered nonzero, as well as zero, depending on whether we consider the parallel influence of both the Coulomb interaction and the $SOC$ or the Coulomb interaction only. The analytical solution for the second case is given in \cite{chakraborty} and we outline the corresponding $LDoS$ in Fig. \ref{graph-coulomb-bezS}. (It is possible to consider nonzero mass in this case but here we suppose, in agreement with (\ref{syst2}), $m_0=0$). In the expression for the analytical solution, $n$ has half-integer values, here we considered $-2.5\leq n\leq 2.5$. The outlined solution, similarly as the solution for the case of the $SOC$ only in Fig. (\ref{LDoS3Dspin}), signalizes the localization of the electrons for $r=0$, but on the contrary to that figure, for higher $r$ and energies close to zero, the $LDoS$ decreases to zero.

For the first case, we use the analog of the numerical method used in \cite{pincak}: we express the solution in the form of an infinite series (Eqs. (A1), (A2) in \cite{pincak}) and substitute it into the system (\ref{syst2}). By the addition of the requirement $\xi_1=\xi+2$, we get a system of recurrence equations for the coefficients $a_k, b_k, c_k, d_k, k\geq0$. From the boundary conditions follows $\alpha=0$ and the same for the coefficients $a_0, b_0, a_1, b_1$. Then, the value of the parameter $\xi$ is determined from the requirement of the nonzero solution of the system for the coefficients $a_2, b_2, c_0, d_0$, i. e. the determinant of the corresponding matrix should be nonzero. In this way, we get
\begin{equation}\label{ksi}\xi=\frac{1}{2}(-5\pm\sqrt{1-4CD-4F+4F^2\pm\sqrt{-CD(1-2F)^2+\kappa^2(C^2-D^2+\kappa^2)}}).\end{equation}
We have to choose one of the 4 possible values of $\xi$; here we chose the value with plus signs on the critical places. Then, the coefficients $a_2, b_2, c_0, d_0$ create the components of the corresponding eigenvector. Now, for $k\geq2$, we can easily calculate the coefficients $a_k, b_k, c_{k-2}, d_{k-2}$ in the chosen limit. The coefficient $\beta$ can be chosen arbitrarily; we chose $\beta=0$.

\begin{figure}[htbp]
\includegraphics[width=170mm]{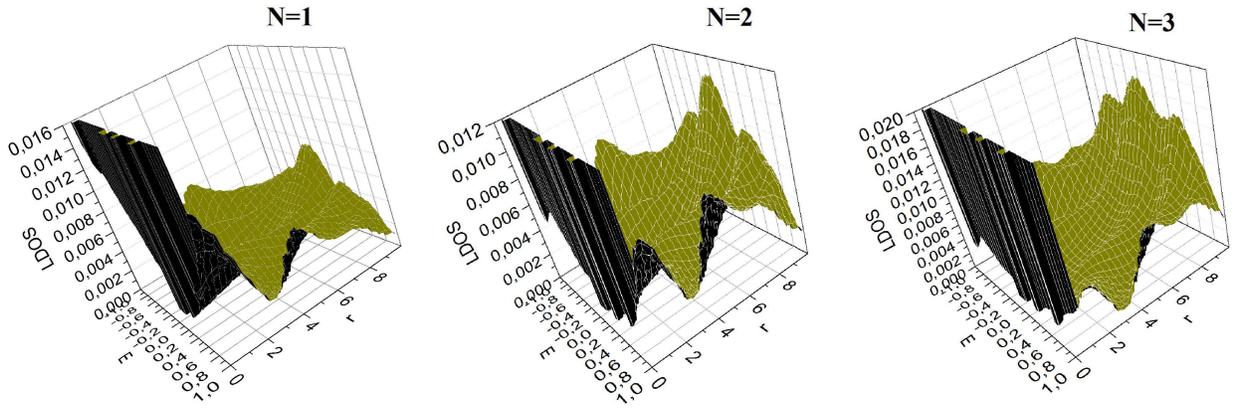}
\caption{Graphs of the $LDoS$
for the graphitic nanocone influenced by the Coulomb interaction (including the influence of the $SOC$) for different
distances $r$ from the tip, $-1\leq n\leq 3$ and for different numbers of the defects.}\label{graph-coulomb}
\end{figure}

The resulting graphs of the $LDoS$ are in Fig. \ref{graph-coulomb}. Here, similarly as in Fig. \ref{LDoS3Dspin}, $-1\leq n\leq 3$. Comparing Figs. \ref{graph-coulomb-bezS} and \ref{graph-coulomb} we see that the same problem appears in both cases: the electrons are localized in the tip for an arbitrary energy for an arbitrary number of the defects and the uniqueness of the peak for zero energy, which corresponded to the mode $n=-1$ and 3 defects in the case of the $SOC$ only (Fig. \ref{LDoS3Dspin}), is distorted by the divergence of the $LDoS$ at all energies for $r=0$.\\

\subsection{Comparison of the results}\

\begin{figure}[htbp]
\includegraphics[width=170mm]{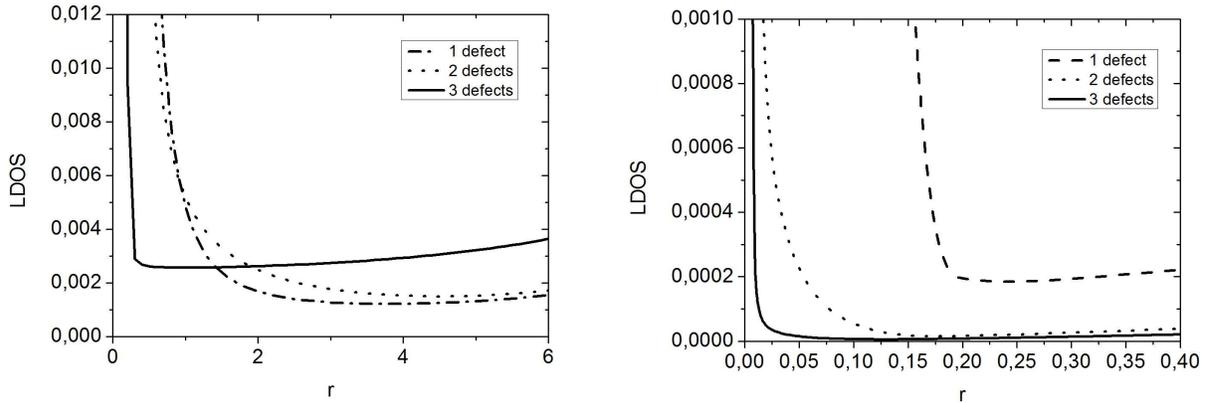}
\caption{Behaviour of the $LDoS$ for zero energy close to $r=0$ for different numbers of defects in the conical tip: influence of the $SOC$ only (left) and the simultaneous influence of the $SOC$ and the Coulomb interaction (right).}\label{LDOSrCS}
\end{figure}

Let us investigate the character of the divergence of the $LDoS$ for $r=0$ closer. In Fig. \ref{LDOSrCS}, we see the dependence of the $LDoS$ on $r$ variable for zero energy in the case of the influence of the $SOC$ only and of the simultaneous influence of the $SOC$ and the Coulomb interaction. We see here that in comparison with the first case, in the second case the decrease of the $LDoS$ close to $r=0$ is much faster and one could suppose that the quadratic integrability of the acquired solution is achieved here. To verify this hypothesis, we have to do the integration of the $LDoS$ close to $r=0$ in all of the outlined cases. This task is still in progress.\\

\begin{figure}[htbp]
\includegraphics[width=170mm]{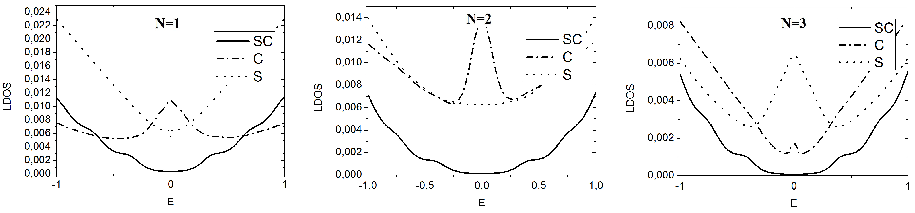}
\caption{Comparison of the influence of the $SOC$ and the Coulomb interaction at the distance $r=1$ from the tip for different numbers of the defects: simultaneous influence of both effects (denoted by SC), the Coulomb interaction only (C) and the $SOC$ only (S).}\label{compare_SC1}
\includegraphics[width=170mm]{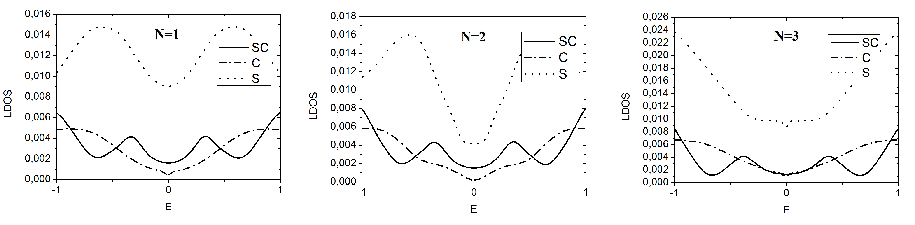}
\caption{The same at the distance $r=5$ from the tip.}\label{compare_SC2}
\end{figure}

In Figs. \ref{compare_SC1} and \ref{compare_SC2}, we see the comparison of all the investigated effects which can influence the electronic structure of the graphitic nanocone. From the graphs follows that at higher distances, the biggest significance has the effect of the $SOC$ and its influence strongly depends on the number of the defects, while the effect of the Coulomb interaction (with or without the $SOC$) nearly does not depend on the number of the defects.

On the other hand, at the lower distance, just after the fast decrease of the $LDoS$ close to $r=0$, the effect of the $SOC$ decreases significantly and this decrease is stronger for the higher number of the defects. The effect of the Coulomb interaction in the presence of the $SOC$ decreases as well for the higher number of the defects and it approaches zero for $E=0$. If the Coulomb interaction occurs without the $SOC$, its influence, thanks to the decrease of other effects, is getting more significant, but significant changes of its magnitude are appreciable only close to zero.\\

\section{Green function approach}\

The results described in the previous sections were acquired by adding different influences to the basic Hamiltonian from \cite{sitenko}, but the presented solutions are distorted by mistakes coming from the used methods of finding the numerical solution. We can verify their validity using different kinds of the Green function method.

The first kind follows from the chemical structure of the molecule and estimated energy of the spin--orbit coupling corresponding to the given atomic site. On this base, we compose the matrix $\mathds{H}(E)$ which describes the energy of the atomic bonds and of the corresponding spin--orbit couplings. Its inverse matrix corresponds to the Green function matrix $\mathds{G}$ from which the $LDoS$ for different atomic sites will be calculated.

The second kind follows from the analytical expression for the Green function corresponding to the pure graphitic nanocone without any other influences \cite{sitenko} from which the Green function corresponding to the spin--orbit coupling will be calculated as a perturbation.

\subsection{Matrix method}\

Using this method, we verify especially the presence of the peak for zero energy close to the conical tip found in section II for the case of 3 defects and mode $n=-1$. It was found using the numerical method of the solution of the Dirac-like equation. The principle of this method consists in the construction of the matrix of the type $(N,N)$, where $N$ is the number of the atoms in the given molecule. In this matrix, the atomic sites are characterized by the diagonal elements and each interatomic bond is characterized by its energy in the non-diagonal element at the appropriate position. For example, for the 2-atomic molecule, the appropriate matrix has the form
\begin{equation}\mathds{H}(E)=\left(\begin{array}{cc}E & t\\t & E\end{array}\right),\end{equation}
where $t$ is the energy corresponding to the interatomic bond. In the case of the consideration of the spin--orbit interaction, we perform this substitution for each matrix element:
\begin{equation}E\quad\rightarrow\quad\left(\begin{array}{cc}E+\triangle(r) & 0\\0 & E-\triangle(r)\end{array}\right)\end{equation}
for the diagonal elements and
\begin{equation}t\quad\rightarrow\quad\left(\begin{array}{cc}t & 0\\0 & t\end{array}\right)\end{equation}
for the non-diagonal elements. Here, $\pm\triangle(r)$ denotes the increase or the decrease, resp. of the energy by the energy corresponding to the spin--orbit interaction depending on the distance from the tip. In the above mentioned case of the 2-atomic molecule it means the transformation of the corresponding matrix in the following way:
\begin{equation}\left(\begin{array}{cc}E & t\\t & E\end{array}\right)\quad\rightarrow\quad
\left(\begin{array}{cccc}E+\triangle_1(r) & 0 & t & 0\\0 & E-\triangle_1(r) & 0 & t\\t & 0 & E+\triangle_2(r) & 0\\0 & t & 0 & E-\triangle_2(r)\end{array}\right)\end{equation}
For the $N$-atomic molecule we create in this way the matrix of the type $(2N,2N)$. Then, the local density of states corresponding to the $m$-th atomic position in this molecule we calculate from the expression
\begin{equation}\label{LDGrM}LDoS(E)=\frac{1}{\pi}{\rm Im}\,(\mathds{G}_{2m-1,2m-1}(E-{\rm i}0)+\mathds{G}_{2m,2m}(E-{\rm i}0)),\end{equation}
where it holds for the Green function matrix $\mathds{G}$
\begin{equation}\mathds{G}(E)=\mathds{H}^{-1}(E).\end{equation}
In this way, we create the corresponding matrices $\mathds{H}(E)$ and we calculate the local density of states for the different atomic positions in the case of the nanocones with different numbers of the defects in the tip. The placement of the defects can differ and in this way, the general result could be influenced.

\begin{figure}[htbp]
{\includegraphics{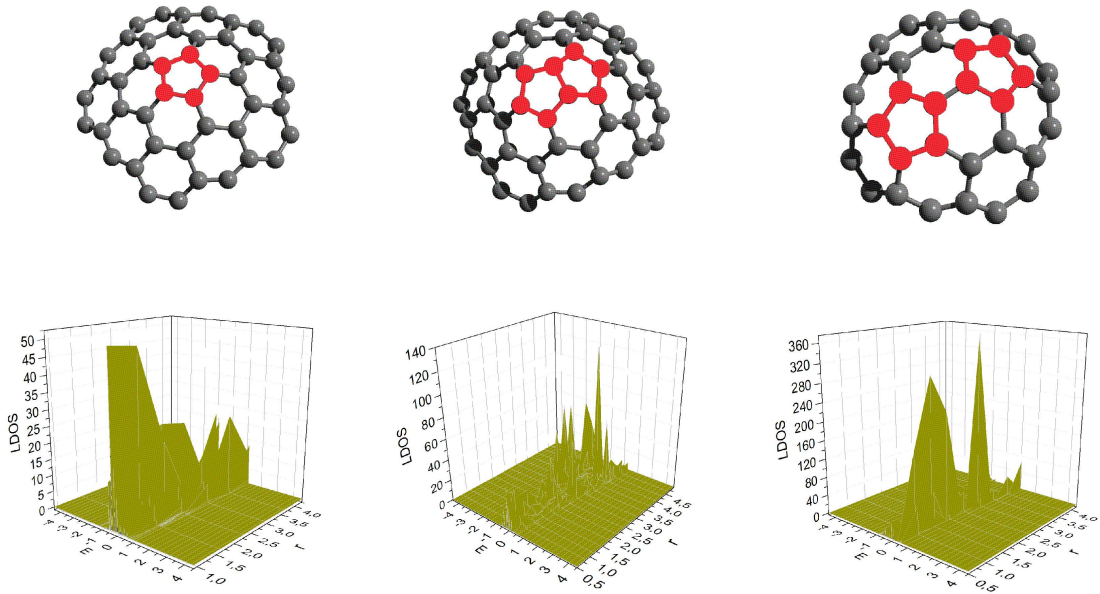}}\caption{Local density of states for nanocones with different numbers and configurations of pentagonal defects in the tip: 1 defect (left), different configurations of 2 defects (middle, right).}\label{12defect}
\end{figure}

\begin{figure}[htbp]
{\includegraphics{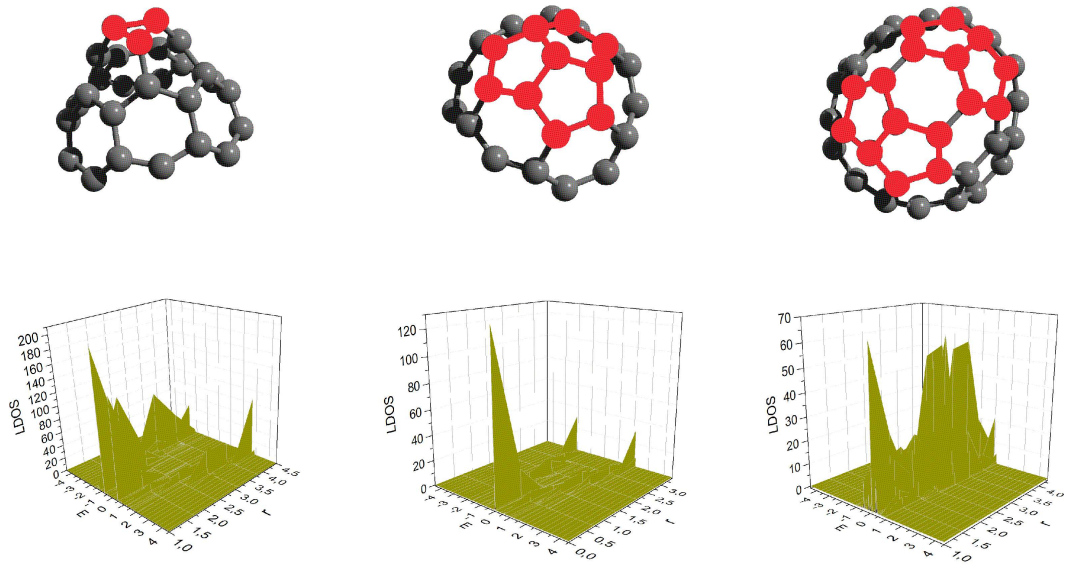}}\caption{Local density of states for nanocones with different numbers and configurations of defects in the tip: triangular defect (left), 3 defects (middle), 4 defects (right). As shown in \cite{PincakMain}, the cases of $n$ pentagonal defects and 1 defect with $6-n$ edges are equivalent, so, the case of triangular defect can be considered as 1 of possible configurations of 3 pentagonal defects.}\label{34defect}
\end{figure}

In Figs. \ref{12defect} and \ref{34defect} we see the local density of states calculated for different configurations of 1, 2, 3 and 4 defects in the tip. In details, the results differ from the results in section II, where all the defects are concentrated in the conical tip, while here they are placed in a limited area around the tip. But the main feature remains the same: In the case of 3 defects, a significant peak around the tip close to zero energy appears. This feature is not so significant for the case of other numbers of the defects - other peaks appear in other distances from the tip and this reduces the significance of possible peaks close to the tip. Moreover, the case of 3 defects is the only case for which we can get an approximate value of $LDoS$ in the tip ($r=0$) - only in this case a configuration with an atom in the tip exists. But still, for other numbers of the defects, we see some of the previously predicted phenomena - for example in Fig. \ref{12defect}, the difference of the character of $LDoS$ for different configurations of 2 defects is in agreement with the predictions from \cite{lammertcrespi}: the configuration in the right part shows much more significant metallic properties than the configuration in the middle part.

In accordance with the results presented in section II for the case of the solution of the Dirac-like equation, in all the results presented in this section the bonds between the carbon atoms are considered the same. In the real nanostructure, the strength of the bonds can differ and we can predict it through the geometric optimization. The dependence $\triangle(r)$ is considered to be linear, so,
\begin{equation}\triangle(r)=\triangle_0\cdot r,\end{equation}
where $\triangle_0=0.8$\,meV\, \cite{steelepei}.

\subsection{Analytical method}\

In this method, we calculate the $LDoS$ using the expression
\begin{equation}\label{LDGreen}LDoS(E)=\frac{1}{\pi}{\rm Im}\,{\rm Tr}\,G(E-{\rm i}0),\end{equation}
where $G$ is the Green function corresponding to the Hamiltonian (\ref{HamSOC}). This expression is formally identical to (\ref{LDGrM}), but here the calculations follow from the analytical expression for the Green function which we can get with the help of the perturbation calculation. It starts from the Green function $G_0$ which corresponds to the Hamiltonian without the influence of the $SOC$ \cite{sitenko}, where the analytical solution is known.

In this calculation, we divide the Hamiltonian (\ref{HamSOC}) as
\begin{equation}
\hat{H}_{s}=\hat{H}_{0s}+\hat{V}_{SOs}
\end{equation}
with
\begin{equation}\hat{H}_{0s}=\hbar v\left(\begin{array}{cc}0 & \left(\partial_r-{\rm i}s\frac{1}{r(1-\eta)}\partial_{\varphi}+\frac{1}{2r}
-\frac{3}{2}\frac{\eta}{(1-\eta)r}\right)\otimes\mathds{1}\\
\left(-\partial_r-{\rm
i}s\frac{1}{r(1-\eta)}\partial_{\varphi}-\frac{1}{2r}
-\frac{3}{2}\frac{\eta}{(1-\eta)r}\right)\otimes\mathds{1}&
0\end{array}\right),\end{equation} and
\begin{equation}\hat{V}_{SOs}=\hbar v\left(\begin{array}{cc}0 & -{\rm i}\frac{\delta\gamma'}{4\gamma R}\sigma_x(\vec{r})-\frac{A_y\sigma_y}{r}\\
{\rm i}\frac{\delta\gamma'}{4\gamma R}\sigma_x(\vec{r})-\frac{A_y\sigma_y}{r}&
0\end{array}\right),\end{equation} here we denote \begin{equation}\mathds{1}=\left(\begin{array}{cc}1 & 0\\
0& 1\end{array}\right).\end{equation} In the perturbation
scheme holds
\[
G(r'',\varphi'',r',\varphi';E)=G_{0}(r'',\varphi'',r',\varphi';E)+G_{1}(r'',\varphi'',r',\varphi';E)+
G_{2}(r'',\varphi'',r',\varphi';E)+...=
\]

\[
=G_{0}(r'',\varphi'',r',\varphi';E)+\int
G_{0}(r'',\varphi'',r,\varphi;E)V_{SOs}(r,\varphi)G_{0}(r,\varphi,r',\varphi';E)r(1-\eta){\rm d}r{\rm d}\varphi+
\]
\begin{equation}
+\int
G_{0}(r'',\varphi'',r_1,\varphi_1;E)V_{SOs}(r_1,\varphi_1)G_{0}(r_1,\varphi_1,r_2,\varphi_2;E)
V_{SOs}(r_2,\varphi_2)G_{0}(r_2,\varphi_2,r',\varphi';E)r_1r_2(1-\eta)^2{\rm d}r_1{\rm d}r_2{\rm d}\varphi_1{\rm d}\varphi_2+...\,\,{\rm etc.}
\end{equation}
Here, $G_{0}(r'',\varphi'',r',\varphi';E)=\langle r'',\varphi''|(H_{0s}-E)^{-1}|r',\varphi'\rangle$ is specified in \cite{sitenko}.
There could be a mismatch thanks to different size of $G_0$ and $V_{SOs}$: the matrix $G_0$ as defined in \cite{sitenko} is of the kind $2\times 2$ while the matrix $V_{SOs}$ is of the kind $4\times 4$. This discrepancy is solved in such a way that $G_0$ in the calculations is considered as $\mathds{1}\otimes G_0$.

We will calculate the approximation of the Green function to the first and to the second order.

\subsubsection{First order approximation}\

In the first order approximation, the Green function has the form
\[\left(G_{1s}\right)_{k,l}=
\frac{\hbar\upsilon\delta}{8\pi}\frac{\gamma'}{4\gamma}\frac{\sqrt{\eta(2-\eta)}}{(1-\eta)^2}\left(\begin{array}{cc}-1 & 0\\ 0& 1\end{array}\right)\sum\limits_{n\in\mathbb{Z}}
\int\limits_0^{+\infty}(a_n^{l1}(r,r')a_{n+1}^{2k}(r',r)-
a_n^{l2}(r,r')a_{n+1}^{1k}(r',r)){\rm d}r+
\]
\begin{equation}\label{GRF}+{\rm i}\frac{\hbar\upsilon\delta ps}{4\pi}\frac{\sqrt{\eta(2-\eta)}}{(1-\eta)^2}\left(\begin{array}{cc}0 & 1\\-1 & 0\end{array}\right)\cdot
\sum\limits_{n\in\mathbb{Z}}\int\limits_0^{\infty}\left(a_n^{l2}(r,r')a_{n}^{1k}(r',r)+
a_n^{l1}(r,r')a_{n}^{2k}(r',r)\right){\rm d}r.
\end{equation}
For the purpose of the calculation of the $LDoS$, we are interested in the trace of the corresponding matrix. Since this trace has the zero value, the first order approximation to the $LDoS$ is zero. It means that the first order approximation is not sufficient to calculate the corrections to the $LDoS$ following from the influence of the $SOC$. The particular elements of the corresponding matrix are nonzero, but it is irrelevant for us for this moment. So we will try to find the result from the approximation to the second order.

\subsubsection{Second order approximation}\

In the second order approximation, the trace of the matrix corresponding to the Green function (which is needed for the definition of the $LDoS$ in (\ref{LDGreen}) ) has the form
\begin{equation}\label{EqG2}{\rm Tr}\,G_{2,s}=-\frac{\hbar^2v^2}{16\pi^2}\frac{\delta^2\eta(2-\eta)}{(1-\eta)^3}\left[\left(\frac{\gamma'}{4\gamma}\right)^2
\sum\limits_{|n_{01}-n_{12}|=1}\int\limits_0^{\infty}
\int\limits_0^{\infty}I_1(r',r_1,r_2){\rm d}r_1{\rm d}r_2-8p^2\sum\limits_{n\in\mathcal{Z}}\int\limits_0^{\infty}
\int\limits_0^{\infty}I_2(r',r_1,r_2){\rm d}r_1{\rm d}r_2\right],\end{equation}
where
{\scriptsize \[I_1(r',r_1,r_2)=a_{n_{01}}^{12}(r_2,r')\left(a_{n_{12}}^{12}(r_1,r_2)a_{n_{01}}^{11}(r',r_1)-
a_{n_{12}}^{11}(r_1,r_2)a_{n_{01}}^{21}(r',r_1)\right)+a_{n_{01}}^{11}(r_2,r')\left(a_{n_{12}}^{21}(r_1,r_2)a_{n_{01}}^{21}(r',r_1)-
a_{n_{12}}^{22}(r_1,r_2)a_{n_{01}}^{11}(r',r_1)\right)+\]}
{\scriptsize \begin{equation}+a_{n_{01}}^{22}(r_2,r')\left(a_{n_{12}}^{12}(r_1,r_2)a_{n_{01}}^{12}(r',r_1)-
a_{n_{12}}^{11}(r_1,r_2)a_{n_{01}}^{22}(r',r_1)\right)+a_{n_{01}}^{21}(r_2,r')\left(a_{n_{12}}^{21}(r_1,r_2)a_{n_{01}}^{22}(r',r_1)-
a_{n_{12}}^{22}(r_1,r_2)a_{n_{01}}^{12}(r',r_1)\right),\end{equation}}
{\scriptsize \[I_2(r',r_1,r_2)=a_{n}^{12}(r_2,r')\left(a_{n}^{12}(r_1,r_2)a_{n}^{11}(r',r_1)+
a_{n}^{11}(r_1,r_2)a_{n}^{21}(r',r_1)\right)+a_{n}^{11}(r_2,r')\left(a_{n}^{22}(r_1,r_2)a_{n}^{11}(r',r_1)+
a_{n}^{21}(r_1,r_2)a_{n}^{21}(r',r_1)\right)+\]}
{\scriptsize \begin{equation}+a_{n}^{22}(r_2,r')\left(a_{n}^{12}(r_1,r_2)a_{n}^{12}(r',r_1)+
a_{n}^{11}(r_1,r_2)a_{n}^{22}(r',r_1)\right)+a_{n}^{21}(r_2,r')\left(a_{n}^{22}(r_1,r_2)a_{n}^{12}(r',r_1)+
a_{n}^{21}(r_1,r_2)a_{n}^{22}(r',r_1)\right).\end{equation}}
Now, we will make the substitution with the help of \cite{sitenko} and we will integrate over $r_1, r_2$ with the help of \cite{abramowitz-stegun}. However, thanks to the appearance of the multiples of 3 coefficients $a_{n}^{kl}(r_i,r_j)$, with the respect to their definition in \cite{sitenko}, the next integration over 3 variables $p_1, p_2, p_3$ is needed. It will be performed numerically and the energy variable will be considered complex in the result.\\

We compare the results acquired by using the Green function approach and with the help of the numerical method used in \cite{pincak}, where the solution for the mode $n=0$ is presented: from (\ref{LDGreen}) follows
\begin{equation}\label{LDOS_2}LDoS_2(E)=\frac{1}{\pi}\,{\rm Im}\,{\rm Tr}\,\left[G_0(E-{\rm i}0)+G_1(E-{\rm i}0)+G_2(E-{\rm i}0)\right],\end{equation}
where the lower index $2$ in the $LDoS_2$ means the mentioned precision. To calculate $G_0$, we exploit the expression (A.6) in \cite{sitenko} for the calculation of ${\rm Tr}\,G_0$ :
\begin{equation}\frac{E}{\pi}\int_0^{\infty}\frac{{\rm d}p\,p}{\hbar^2v^2p^2-E^2}\sum\limits_{n\in\mathcal{Z}}
[J_{2sn-1}(pr)J_{2sn-1}(pr')+J_{2sn}(pr)J_{2sn}(pr')],\end{equation}
where $r\rightarrow r'$. In the sum, we use only the term $J_{2sn}(pr)J_{2sn}(pr')$ for $n=0$. Tr\,$G_1$ is zero for arbitrary $n$, as follows from our calculations for the first order approximation. We calculate Tr\,$G_2$ using (\ref{EqG2}).

\begin{figure}[htbp]
\centering
\includegraphics[width=55mm]{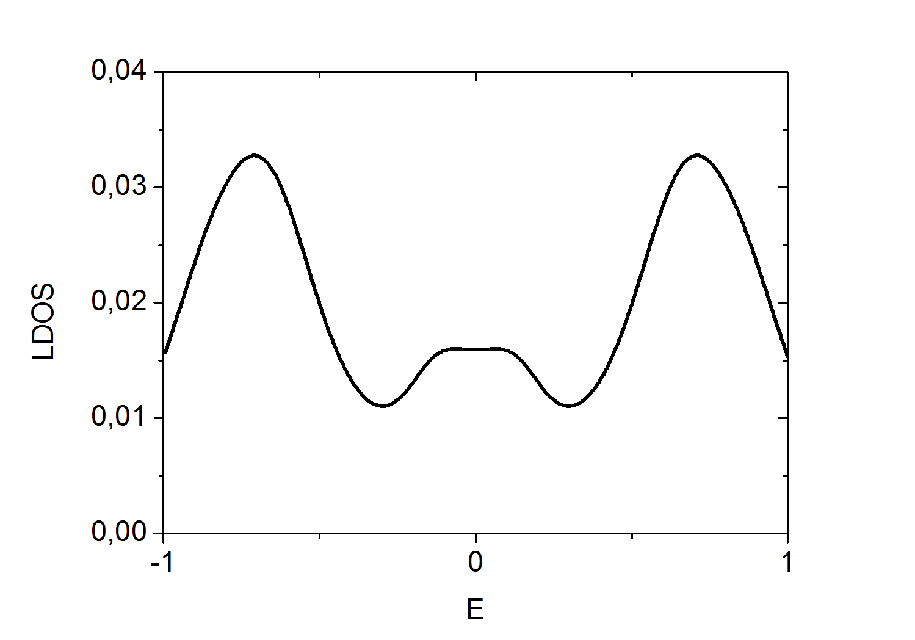}
\includegraphics[width=55mm]{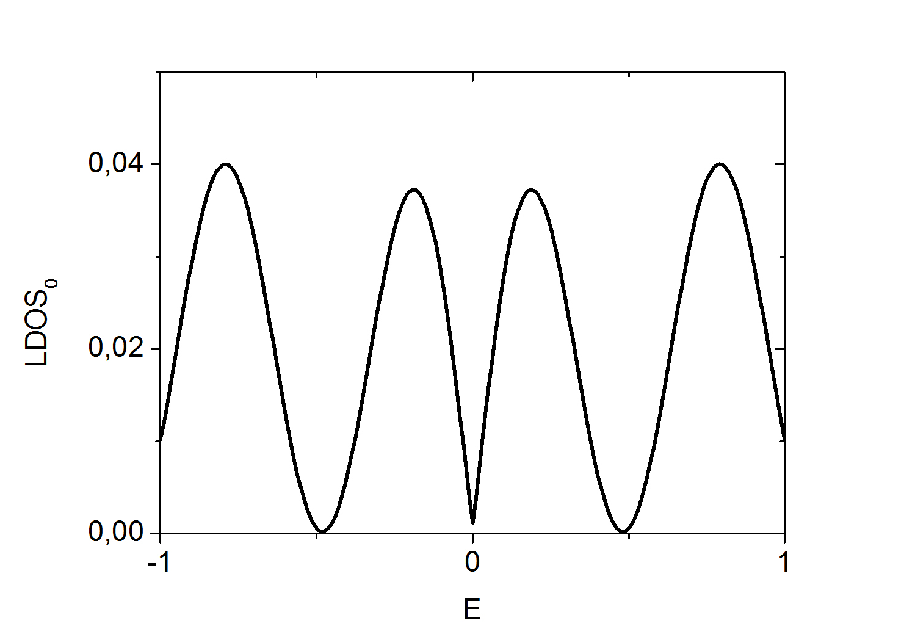}
\includegraphics[width=55mm]{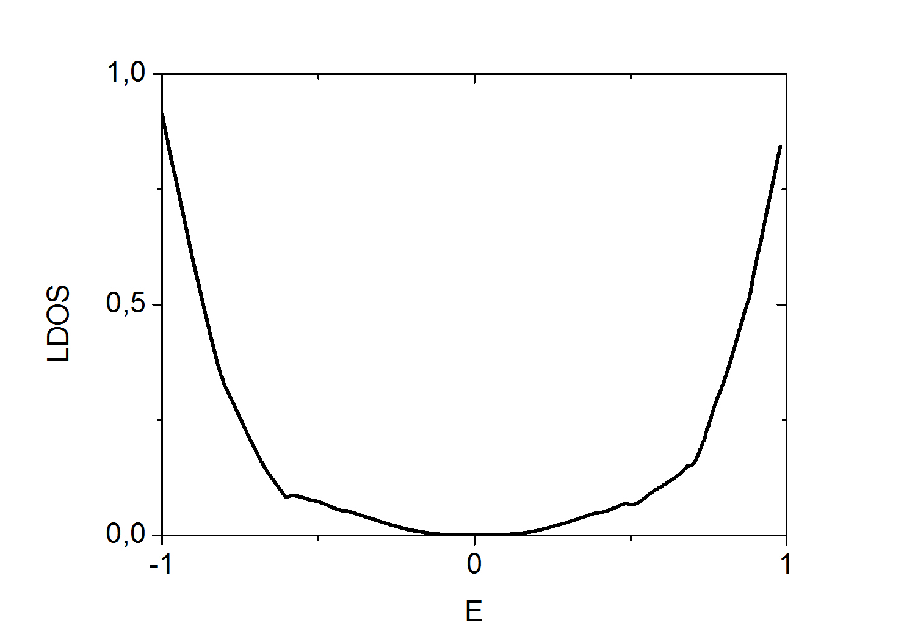}
\caption{$LDoS$ calculated using the numerical method in \cite{pincak} (left - see also the right part of Fig. 3 in \cite{pincak}), zeroth order approximation calculated using the relations in \cite{sitenko} (middle) and second order approximation calculated using (\ref{EqG2}) (right), performed for 3 defects and the values are $n=0$, $r'=5$.}\label{LDOS_Green_Fg}
\end{figure}

If the second order approximation is sufficient, then the graph in the left part of Fig. \ref{LDOS_Green_Fg} should be given by the sum of the graphs in the middle and in the right part. But at first glance, the graphs in the middle and in the right part have different scales, so their sum cannot approach the graph in the left part. The reason is that no normalization was considered during the calculation of the Green function approach. So, using (\ref{LDOS_2}), we try to evaluate the expression
\begin{equation}\label{approx}\frac{1}{\pi}\,{\rm Im}\,{\rm Tr}\,\left[G_0(E-{\rm i}0)+\mathcal{N}G_2(E-{\rm i}0)\right],\end{equation}
where $\mathcal{N}$ is the normalization constant.

\begin{figure}[htbp]
\centering
\includegraphics[width=80mm]{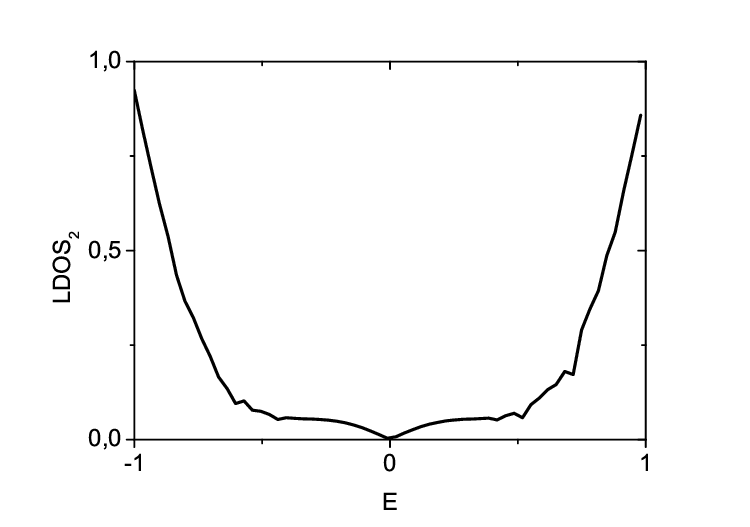}
\includegraphics[width=80mm]{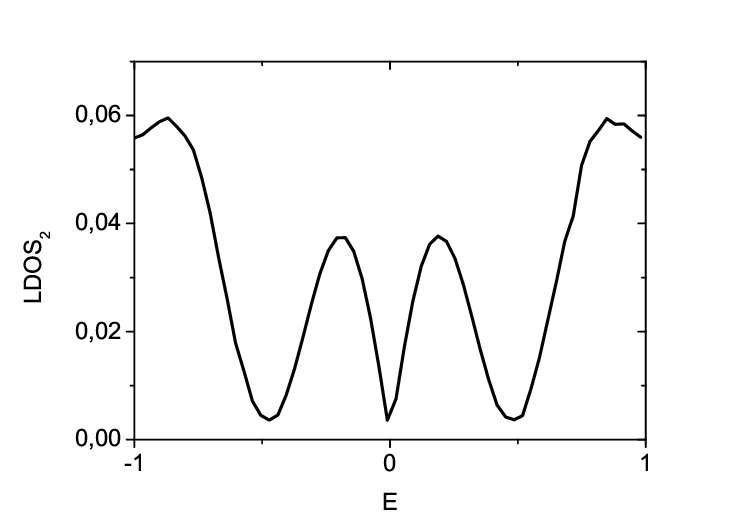}
\caption{Expression (\ref{approx}) evaluated using 2 different values of $\mathcal{N}$: $\mathcal{N}=1$ (left) and $\mathcal{N}=1/20$ (right).}\label{LDOS_approx_Fg}
\end{figure}

In Fig. \ref{LDOS_approx_Fg}, we see that (\ref{approx}) will coincide with the left part of Fig. \ref{LDOS_Green_Fg} much better for the case $\mathcal{N}=1/20$. But still it is not complete coincidence, especially for the energy values close to zero. So it means that the next order of the Green approach is needed.

\subsubsection{Case of the finite size}\

In the case of the finite size of the conical structure, the energy levels would be quantized and the Green function approach would be modified for the case of the discrete spectrum. The discrete energy levels would be calculated on the base of an additional requirement on the wave function which should have the zero value on the border. We use the simplest case of the nanocone without the $SOC$ presented in \cite{sitenko}. This case provides the solution of the form (see (49) in \cite{sitenko})
\begin{equation}\psi(r)=\frac{1}{2\sqrt{\pi(1-\eta)}}\left(\begin{array}{c}J_{\nu}(\frac{|E|}{\hbar v}r)\\{\rm sign}\,E J_{\nu+1}(\frac{|E|}{\hbar v}r)\end{array}\right).\end{equation}
Particular parameters have this form:
\begin{equation}\nu=\frac{sn-\eta}{1-\eta},\hspace{1cm}\hbar v=\frac{3}{2}t\end{equation}
where $t$ is the energy of the C-C bond (about 3 eV). So we require that one of the components of the wave function has the zero value on the border:
\begin{equation}J_{\nu}(r_{max}\cdot\frac{2}{3}|E'|)=0,\,\,{\rm resp.}\,\,J_{\nu+1}(r_{max}\cdot\frac{2}{3}|E'|)=0,\end{equation}
where $E'$ and $r_{max}$ are measured in the units of $t$ and the length of C-C bond, respectively. To demonstrate the character of the resulting energy levels, we will investigate the cases of $1$ and $3$ defects, i.e. $\eta=1/6$ and $\eta=1/2$ and make the choice of the parameters $s=1, n=2$. Next, we take the value $r_{max}=20$ as the border value of $r$. Then, for the particular cases, the low-energy spectrum contains these values:

\begin{description}
\item[-] $1$ defect: $\nu=2.5$ and $E'=0,\,\,\pm0.432259,\,\,\pm0.682126,\,\,\pm0.924221,\,\,\pm1.1636,\,\,\pm1.40168,\,\,\pm1.63904,\,\,\pm1.87596$ or\\
\hspace*{0.7cm}         $\nu=3.5$ and $E'=0,\,\,\pm0.524095,\,\,\pm0.781284,\,\,\pm1.02735,\,\,\pm1.26927,\,\,\pm1.50914,\,\,\pm1.74782,\,\,\pm1.98576$
\item[-] $3$ defects: $\nu=3$ and $E'=0,\,\,\pm0.478512,\,\,\pm0.732077,\,\,\pm0.97614,\,\,\pm1.21676,\,\,\pm1.45571,\,\,\pm1.6937,\,\,\pm1.93111$ or\\
\hspace*{0.7cm}         $\nu=4$ and $E'=0,\,\,\pm0.569126,\,\,\pm0.829853,\,\,\pm1.07794,\,\,\pm1.3212,\,\,\pm1.56202,\,\,\pm1.80143$
\end{description}

We see that the dependence of the energy levels on the number of the defects in the tip is not very strong. To calculate the resulting $LDoS$, the Green function approach for the discrete spectrum would be performed with the help of the summation over other modes $n$.

\section{Conclusion}\

The boundary effects influencing the electronic structure of the graphitic nanocone were investigated and we used mostly numerical methods for the corresponding calculations. We wanted to find a quadratically integrable solution which would preserve the main features of the solution found in \cite{pincak}, especially the isolated peak for $r=0$ and $E=0$ in the case of 3 defects which could have a potential use for the construction of the probing tip in the atomic force microscopy. We achieved this goal only partially - the found solution for the case of the simultaneous influence of the Coulomb interaction and the $SOC$ in Fig. \ref{graph-coulomb} (as well as the analytical solution for the case of the Coulomb interaction only from \cite{chakraborty} sketched in Fig. \ref{graph-coulomb-bezS}) seems to be quadratically integrable, but the mentioned isolated peak for the case of 3 defects disappeared. We also tried to find the possible corrections caused by the considerable rise of the mass of the electron bounded on the molecular surface close to $r=0$ (coming from the rapid curvature), but the resulting changes of the $LDoS$ connected with this effect seem to be negligible (regardless we use the numerical or the analytical solution).

For the case of the spin--orbit interaction, the numerical results were verified using the matrix form of Green function method for the different numbers and configurations of the defects. It was shown that although some small disagreements in the results were presented, the main character of the $LDoS$ remained the same, especially for the case of 3 defects and the zero energy peak close to the tip. We can also do a comparison with the results in \cite{ulloa}, where the calculations were performed for the case of the nanocone with 0 (nanodisk), 1 and 2 pentagonal defects in the tip. The investigated nanostructures contain very high number of carbon atoms - the nanostructures with the number of atoms about 250 and 5000 were chosen. The results also show an appearance of a zero energy peak in $LDoS$ close to the tip. For the case of 5000 atoms, the height of the peak does not depend on the number of the defects, for the case of 250, the height of this peak increases a little with the number of the defects. So, we can suppose that for our case of tens of atoms, the results would confirm our calculations - the peak would be much more significant for higher number of defects, especially the case of 3 defects.

The results coming from the analytical form of the Green function approach for the continuous spectrum (infinite size of the nanocone) don't correlate sufficiently with the results of the numerical approach in \cite{pincak}. This is given by more reasons: first, we performed the calculations of the difficult integrals numerically with the help of the program Mathematica, so the precision of the acquired results is limited. Secondly, we did the calculations to the second order only. Maybe we should use also the third order approach.

In further calculations, we could also consider the electronic structure of the double graphitic nanocone. The corresponding Hamiltonian would differ by the presence of a new contribution which would be constant for all range of $r$ excluding a small interval close to the tip, where a strong influence of the interaction with the neighboring atomic orbitals caused by the curvature would increase the resulting effect.

Although up to now, no effective method for the fabrication of the graphene nanocones was found, an evidence of occurrence of these materials was described in \cite{a1}. Moreover, next theoretical calculations and molecular simulations were performed \cite{a2, a3}. These results promise practical applications of the graphitic nanocones in close future. New calculations related to the spin--orbital coupling in graphene \cite{a4} as well as the theoretical and experimental investigations of graphene monolayer and bilayer from the last days \cite{a5, a6} contribute to this goal.\\

ACKNOWLEDGEMENTS --- The work was supported by the Science and Technology Assistance Agency under Contract No. APVV-0171-10, VEGA Grant No. 2/0037/13 and Ministry
of Education Agency for Structural Funds of EU in the frame of project
 26220120021, 26220120033 and 26110230061. R. Pincak would like to thank the
 TH division at CERN for hospitality.

\end{document}